# Effective lattice Hamiltonian for monolayer tin disulphide: tailoring electronic structure with electric and magnetic fields

Jin Yu[1,2], Edo van Veen[2], Mikhail I. Katsnelson[2] and Shengjun Yuan[3,1,2]

1 Beijing Computational Science Research Center, Beijing 100094, China

2 Theory of Condensed Matter, Radboud University, Nijmegen 6525AJ, The Netherlands

3 School of Physics and Technology, Wuhan University, Wuhan 430072, China

**Abstract**

The electronic properties of monolayer tin dulsulphide (ML-$SnS_2$), a recently synthesized metal dichalcogenide, are studied by a combination of first-principles calculations and tight-binding (TB) approximation. An effective lattice Hamiltonian based on six hybrid *sp*-like orbitals with trigonal rotation symmetry are proposed to calculate the band structure and density of states for ML-$SnS_2$, which demonstrates good quantitative agreement with relativistic density functional theory calculations in a wide energy range. We show that the proposed TB model can be easily applied to the case of an external electric field, yielding results consistent with those obtained from full Hamiltonian results. In the presence of a perpendicular magnetic field, highly degenerate equidistant Landau levels are obtained, showing typical two-dimensional electron gas behavior. Thus, the proposed TB model provides a simple new way in describing novel properties in ML-$SnS_2$.

**Introduction**

Two-dimensional (2D) materials have been attracting extensive attention as the key building blocks for next generation electronic, photonic and optoelectronic systems because of their ultrathin atomic structure and novel physical properties. [1] Special interest has been put on layered metal dichalcogenides, which exhibit both fundamentally and technologically interesting properties. [2-5] As an important member of the layered metal disulfide family, monolayer tin disulfide (ML-$SnS_2$) has drawn considerable attention recently due to its low production cost, high chemical stability, excellent photosensitivity and superior photoelectric properties. [6-9] Compared to other metal dichalcogenides, tin (Sn) is lighter in mass than conventional transition metals, and ML-SnS$_2$ has weak spin orbital coupling, which can be attributed to its outermost electron being dominated by the *s* orbital. Like most 2D materials reported, high-quality few-layer ML-SnS$_2$ can be obtained by chemical vapor deposition [10, 11] or mechanical exfoliation from layered bulk crystals [12, 13]. ML-SnS$_2$ is reported to have a visible-light band gap around 2.2 eV, which offers possibilities in solar cells design and visible-light water splitting manipulation. [14-16] Moreover, its high ratio area enables it with high reversible capacity as anode material in lithium and sodium ion batteries. [17-20] Because of its relatively high carrier mobility and on-off current ratio, ML-SnS$_2$ has the advantage of suppressing drain to source tunneling for short channels, rendering it a promising candidate in field-effect transistors, integrated logic circuits and photodetectors. [9, 12]

The electronic properties of ML-SnS$_2$ have been studied by first-principles calculations. [21-25] However, these methods usually have a high computational cost, and can only consider a limited number of atoms. This is not enough to describe the properties of realistic materials at large scales and heterojunctions with a super lattice like ML-SnS$_2$/WSe$_2$ [26]. Alternatively, the method of model Hamiltonians paves a way to address this problem - it is less transferable, but very efficient and flexible. Several tight-binding (TB) models have shown great success in capturing the relevant electronic states in other 2D materials like graphene [27, 28] and its derivatives and heterostructures [29, 30], black phosphorus [31], monolayer antimony [32] and transition metal dichalcogenides [33-35]. Therefore, it is useful to construct an effective Hamiltonian for ML-SnS$_2$ as well, to further study its electronic properties.

In this paper, we present a suitable model Hamiltonian governing the low-energy band structure of ML-SnS$_2$ without spin-orbit coupling and show that its electronic properties can be tuned by applying a perpendicular electric field or magnetic field. After analyzing the orbital character of the electronic states at the relevant high-symmetry points, we build the TB model consisting of six orbitals. These mainly consist of sulfur (S) 3*p* orbital and tin (Sn) 4*s* orbital hybrids. We calculate the band structure of the TB model, and compare it to density functional theory calculations. Next, we turn to the effect of an external electric field, which leads to shifts in the chemical potentials of three sublayers consisting of one Sn and two S layers, resulting in a splitting of low-energy bands originating from the *p* orbitals of S atoms. We also show that the electronic properties of ML-SnS$_2$ can be modulated by applying a perpendicular magnetic field which induces highly degenerated Landau levels. Generally, our proposed TB model captures the dominant contribution from Sn *s* and S *p* orbitals in the low energy region, and thus it can be considered as a starting point to study the electronic states of nanoribbons, defects, impurities and multi-layers in ML-SnS$_2$ at large scales.

**Computational methods**

In constructing a reliable TB model for semiconducting SnS$_2$, we are guided by first-principles calculations that will provide the reference on which to calibrate the effective Hamiltonian. Equilibrium structural parameters and reference electronic bands were obtained at the density functional theory (DFT) level using VASP code. [36, 37] The generalized gradient approximation [38] was used in combination with the projected augmented-wave method [39]. The vacuum distance of ML-SnS$_2$ between two adjacent images was set to be at least 1.2 nm. The plane wave cutoff energy was set to 280 eV. The Brillouin zone sampling was done using a 15 × 15 × 1 Monkhorst-Pack grid for relaxation calculations and a 35 k-point sampling was used for the static calculations. All the atoms in the unit cell were fully relaxed until the force on each atom was less than 0.01 eV/A. Electronic minimization was performed with a tolerance of 10$^{-5}$ eV. The construction of the Wannier functions and TB parametrization of the DFT Hamiltonian were done with the WANNIER90 code [40]. The electronic density of states with external magnetic field was calculated from the solution of the time-dependent Schrödinger equation within the framework of the tight-binding propagation method,[41, 42] which is an efficient numerical tool in large-scale calculations of realistic systems with more than millions of atoms.

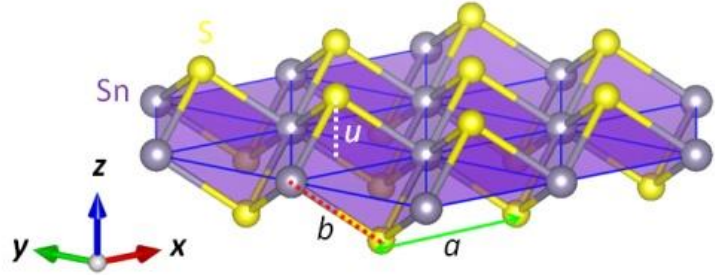

**Figure 1** Atomic structure of ML-SnS$_2$. The single layer is built up from two trigonal pyramids. The structural parameters are defined in the text.

**Results**

Unlike monolayer MoS$_2$, which has a 1H-phase ground state structure, ML-SnS$_2$ prefers a 1T-phase structure, which is depicted in Figure 1. It has a hexagonal unit cell with space group $D_{3d}^{-3}$. The basic unit block is composed of one layer of Sn atoms surrounded by two layers of S atoms in such a way that each Sn atom is coordinated by six S atoms in two pyramidal geometries and each S atom is coordinated by three Sn atoms, where strong covalent bonding exists in plane but weak van der Waals interaction dominates in the out of plane direction. Following the notation of MoS$_2$, we denote $a$ as the distance between nearest-neighbor in-plane Sn-Sn distances, $b$ as the nearest-neighbor Sn-S distances, and $u$ as the distance between the S and Sn planes. We set $a$ = 3.703 Å, $b$ = 2.674 Å and $u$ = 1.606 Å, respectively.

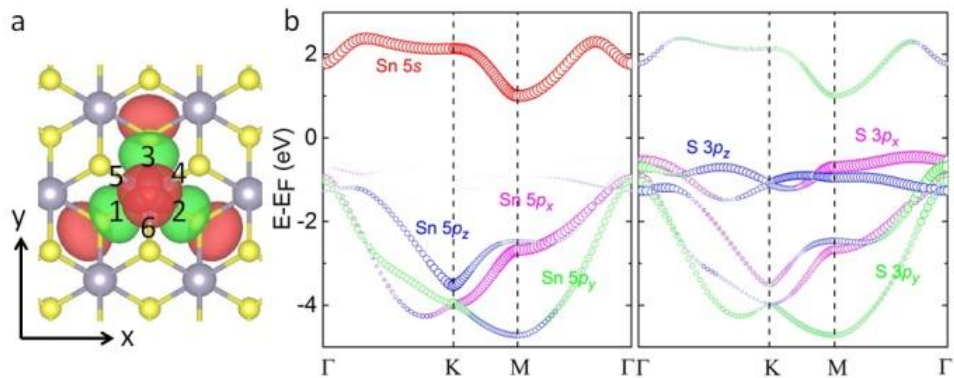

**Figure 2** Orbital characters of ML-SnS$_2$. (a) Wannier orbitals of ML-SnS$_2$ corresponding to the basis of the TB Hamiltonian presented in this work. For clarity, orbitals are shown for one S-atom sublattice with three orbitals per atom.

The orbitals in the second sublattice are symmetric with respect to the inversion center. (b) Orbital decomposed band structure of Sn and S in ML-SnS$_2$, with the corresponding *s, p$_x$, p$_y$* and *p$_z$* orbitals being red, pink, green and blue, respectively. The size of the symbol represents the orbital weight.

The in-plane Brillouin zone in a hexagonal unit cell is thus characterized by the high-symmetry points Γ = $2\pi/a(0, 0)$, M = $2\pi/a(0, 1/\sqrt{3})$ and K = $2\pi/a(1/3, 1/\sqrt{3})$. Since a special role in the electronic properties of ML-SnS$_2$ is played by the electronic states at the conduction band minimum (CBM) and the valence band maximum (VBM), our results mainly focus on the block of bands containing the first five valence bands and the first conduction band in the energy window from -5 eV to 3 eV. The orbital decomposed band structure from DFT calculations for Sn and S are shown in Figure 2b. An accurate description of the conduction band and valence bands in this energy window involves at least the *s* orbital of Sn and the *p* orbitals of S atoms. Detailed analysis of the band structure shows that the VBM is located slightly away from the Γ point, where the first and second valence bands are doubly degenerate and the corresponding states are composed of the *p$_x$* and *p$_y$* orbitals from the top and bottom S atoms, which are symmetric with respect to the inversion center. At a much lower energy level, there is another doubly degenerate band, with contributions from the *p$_x$* and *p$_y$* orbitals from both the Sn and S atoms. The CBM is located at the M point, where an orbital decomposition of the corresponding wave function yields $|\varphi^{CB}(M)\rangle = 0.133|s\rangle_S + 0.717|s\rangle_{Sn} + 0.634|p_y\rangle_S + 0.259|p_z\rangle_S$. The indirect gap between the VBM and CBM points is estimated to be ~1.16 eV.

As the hole and electron states are symmetry inequivalent, we can construct a non-trivial TB model for ML-SnS$_2$. Considering that the valence and conduction bands are dominated by hybrid *sp-like* orbitals and that they are separated from other states, it is possible to provide an accurate description of those states in terms of a tractable TB model in the low-energy region. The parametrization procedure used in our work is based on the formalism of maximally localized Wannier functions (WF). [43-45] where, the cell periodic part $u_{n\mathbf{k}}^H(\mathbf{r})$ of the Bloch function

$$\varphi_{n\mathbf{k}}^H(\mathbf{r}) = u_{n\mathbf{k}}^H(\mathbf{r})e^{i\mathbf{k}\cdot\mathbf{r}}, \quad (1)$$

represents the eigenfunctions of the first-principles Hamiltonian $H^H(\mathbf{k})$. And it transforms according to

$$u_{n\mathbf{k}}^W(\mathbf{r}) = \sum_m U_{mn}^{\mathbf{k}} u_{m\mathbf{k}}^H(\mathbf{r}), \quad (2)$$

where $n$ is the band index and $\mathbf{k}$ is the crystal momentum, $U_{mn}^{\mathbf{k}}$ is a unitary matrix chosen to minimize the spread of the Wannier orbitals

$$w_{n\mathbf{R}_i}(\mathbf{r}) = \frac{1}{N_k} \sum_{\mathbf{k}} e^{-i\mathbf{k}\cdot\mathbf{R}_i} \varphi_{n\mathbf{k}}^W(\mathbf{r}), \quad (3)$$

which is centered at $\mathbf{R}_i$. A real-space distribution of the WFs obtained for ML-SnS$_2$ is shown in Figure 2a, where a combination of three hybrid *sp*-like orbitals are represented around each S atom, giving rise to six WFs per unit cell on two sublattices. These three orbitals are all equivalent and have a rotation symmetry of $2\pi/3$, which is helpful in reducing the parameter numbers.

Then, the resulting nonrelativistic TB model is given by an effective full Hamiltonian,

$$H_0 = \sum_{mn} \sum_{ij} t_{ij}^{mn} c_{im}^\dagger c_{jn}, \quad (4)$$

where $t_{ij}^{mn}$ is the effective hopping parameter describing the interaction between $m$ and $n$ orbitals residing at atoms $i$ and $j$, respectively. Moreover, $c_{im}^\dagger$ ($c_{jn}$) is the creation (annihilation) operator of electrons at atom $i$ ($j$) and orbital $m$ ($n$). In order to identify the most relevant hopping processes, we first discard hoppings with an interatomic distance larger than 8.35 Å. To make the model even more simple, we ignore hopping parameters with amplitudes $|t_i| < 10$ meV. This choice of cut-offs ensures a model that is simple, but accurate enough for further calculations. These hopping parameters are further re-optimized through the least square method to get close to the full Hamiltonian model. The remaining orbitals and the relevant hopping parameters are schematically shown in Figure 3 and Table 1. If without special notation, all our TB results are based on this simple Hamiltonian model.

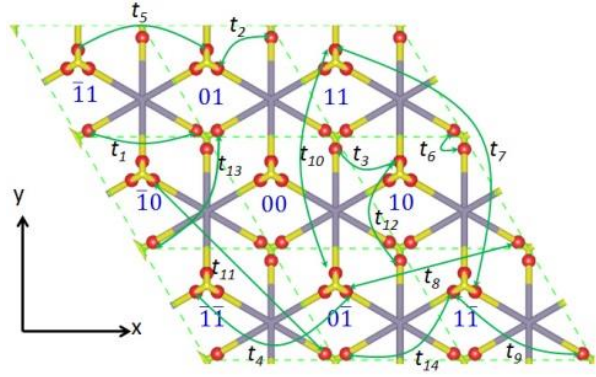

**Figure 3** Schematic representation of the crystal structure (top view) and relevant hopping parameters ($t_i$) involved in the simple TB model of ML-SnS$_2$. Interacting hopping centers are depicted by red balls, corresponding to the negative part of the Wannier orbitals (cf. Figure 2). The hopping amplitudes are given in Table 1. Blue labels mark relative unit cell coordinates.

**Table 1** Hopping amplitudes $t_i$ (in eV) assigned to the TB Hamiltonian of ML-SnS$_2$. $d$ denotes the distance between the lattice sites on which the interacting orbitals are centered. $N_c$ is the corresponding coordination number. The hoppings are schematically shown in Figure 3.

| i | $t_i$(eV) | d(Å) | $N_c$ | i | $t_i$(eV) | d(Å) | $N_c$ | i | $t_i$(eV) | d(Å) | $N_c$ |
|---|---|---|---|---|---|---|---|---|---|---|---|
| 1 | -0.44 | 3.70 | 2 | 6 | -0.02 | 0 | 1 | 11 | 0.05 | 8.35 | 2 |
| 2 | -0.42 | 3.86 | 2 | 7 | 0.02 | 7.41 | 2 | 12 | 0.09 | 3.86 | 1 |
| 3 | -0.36 | 3.86 | 2 | 8 | -0.02 | 6.50 | 2 | 13 | 0.28 | 3.70 | 4 |
| 4 | 0.24 | 3.70 | 1 | 9 | -0.02 | 5.35 | 2 | 14 | 1.15 | 5.35 | 1 |
| 5 | -0.07 | 3.70 | 2 | 10 | 0.03 | 6.41 | 2 | | | | |

In Equation (4), our TB model is defined in terms of a 6 × 6 Hamiltonian which can now be explicitly solved to get eigenvalues and eigenvectors. Due to inversion symmetry of the atomic structure, the reciprocal space Hamiltonian matrix can be further simplified as

$$H(\mathbf{k}) = \begin{pmatrix} E(\mathbf{k}) & T(\mathbf{k}) \\ T^\dagger(\mathbf{k}) & E(\mathbf{k}_r) \end{pmatrix}, \quad (5)$$

where $E(\mathbf{k})$ and $T(\mathbf{k})$ are $3 \times 3$ matrices describing the intrasublattice and intersublattice interactions, respectively. Considering the trigonal rotation symmetry, the corresponding matrices have the simplified forms

$$E(\mathbf{k}) = \begin{pmatrix} A(\mathbf{k}) & B(\mathbf{k}) & B^*(\bar{\bar{\mathbf{k}}}) \\ B^*(\mathbf{k}) & A(\bar{\mathbf{k}}) & B(\bar{\mathbf{k}}) \\ B(\bar{\bar{\mathbf{k}}}) & B^*(\bar{\mathbf{k}}) & A(\bar{\bar{\mathbf{k}}}) \end{pmatrix}, (6)$$

and

$$T(\mathbf{k}) = \begin{pmatrix} C(\mathbf{k}) & D(\mathbf{k}) & C(\bar{\bar{\mathbf{k}}}) \\ D(\bar{\mathbf{k}}) & C(\mathbf{k}) & C(\bar{\mathbf{k}}) \\ C(\bar{\bar{\mathbf{k}}}) & C(\bar{\mathbf{k}}) & D(\bar{\mathbf{k}}) \end{pmatrix}. (7)$$

In equation (6) and (7), $\bar{\mathbf{k}}$ ($\bar{\bar{\mathbf{k}}}$) is the $\mathbf{k}$ vector rotated by $2\pi/3$ ($4\pi/3$), whereas the subscript $r$ of $\mathbf{k}$ in equation (5) indicates rotation in the opposite direction, equivalent to the inversion symmetry operation. The matrix elements appearing in equation (6) and (7) read

$A(\mathbf{k}) = 2t_{13} \cos\left(\frac{1}{2}k_x a + \frac{\sqrt{3}}{2}k_y a\right) + 2t_{13} \cos(k_x a) + 2t_5 \cos(-\frac{1}{2}k_x a + \frac{\sqrt{3}}{2}k_y a) + 2t_{10} \cos(\frac{3}{2}k_x a + \frac{\sqrt{3}}{2}k_y a)$, (8)

$B(\mathbf{k}) = t_6 + t_1 e^{ik_x a} + t_4 e^{-ik_x a} + t_7 e^{2ik_x a}$, (9)

$C(\mathbf{k}) = t_3 e^{i\frac{\sqrt{3}}{3}k_y a} + t_3 e^{-i(-\frac{1}{2}k_x a + \frac{\sqrt{3}}{6}k_y a)} + t_{14} e^{i(\frac{1}{2}k_x a + \frac{\sqrt{3}}{3}k_y a)} + t_8 e^{i(\frac{1}{2}k_x a + \frac{5\sqrt{3}}{6}k_y a)} + t_8 e^{i(\frac{3}{2}k_x a - \frac{\sqrt{3}}{6}k_y a)} + t_{12} e^{-i(\frac{1}{2}k_x a + \frac{\sqrt{3}}{6}k_y a)} + t_{11} e^{i(\frac{3}{2}k_x a + \frac{5\sqrt{3}}{6}k_y a)} + t_{11} e^{i(2k_x a + \frac{\sqrt{3}}{3}k_y a)}$, (10)

$D(\mathbf{k}) = t_2 e^{i\frac{\sqrt{3}}{3}k_y a} + t_9 e^{-i(k_x a - \frac{\sqrt{3}}{3}k_y a)} + t_9 e^{i(k_x a + \frac{\sqrt{3}}{3}k_y a)}$, (11)

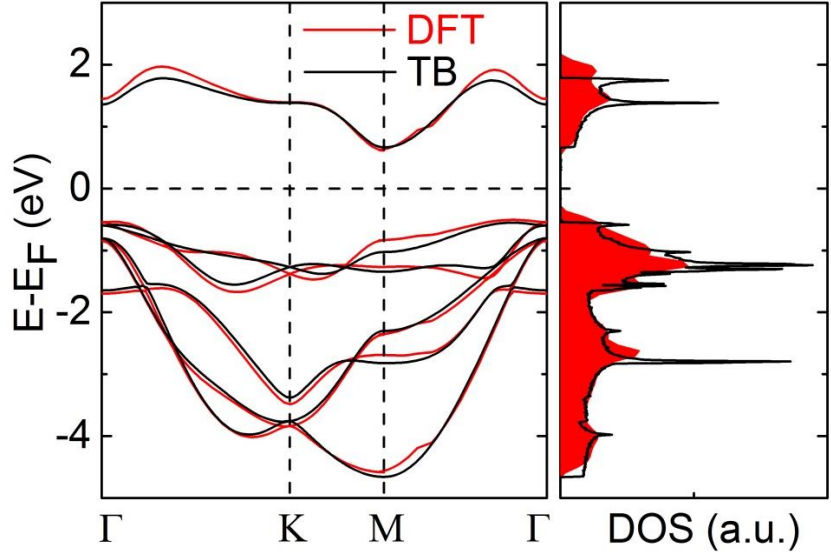

**Figure 4** Band structure (left) and density of states (right) calculated for ML-SnS$_2$ using the DFT and simple TB model presented in this work.

A comparison of the band structure and DOS for the DFT and TB models is shown in Figure 4. The TB model agrees in a qualitative way with the original first-principles calculations. In particular, it shows an indirect gap between the Γ and M points, and a secondary direct band gap for the valence and conduction bands lying at the M point. As we ignore some hopping terms other than present *sp*-like orbitals, the total DOS from DFT calculations has a slightly wider energy range than the TB model.

The agreement of the DFT results in the low-energy region can be further quantified by the band gaps and carrier effective masses, which are accurately reproduced by the proposed TB model as shown in Table 2. The indirect and secondary direct band gap in our TB model are calculated to be 1.26 eV and 1.69 eV, respectively, which are slightly overestimated mainly because our effective model Hamiltonian does not provide a good description of the first valence band, especially for the holes around the M point. However, electrons at M point are well described as can be seen from the anisotropic effective mass of electrons at the CBM. There, the effective masses of electrons along the x and y direction are calculated to be 0.31 $m_0$ and 0.77 $m_0$ from DFT, and the corresponding values in our TB model are 0.33 $m_0$ and 0.72 $m_0$, respectively.

**Table 2** Indirect (ΓM) and direct (MM/ΓΓ) band gaps, $E_g$ (in eV), as well as effective masses $m$ (in units of the free electron mass $m_0$) calculated for holes and electrons in ML-SnS$_2$ at relevant high-symmetry points of the Brillouin zone using the DFT and TB model present in this work.

| Method | Energy gap (eV) | | | Holes ($m_0$) | | Electrons ($m_0$) | | | |
|---|---|---|---|---|---|---|---|---|---|
| | $E_g^{\Gamma M}$ | $E_g^{\Gamma\Gamma}$ | $E_g^{MM}$ | $m^{\Gamma K}$ | $m^{M\Gamma}$ | $m^{\Gamma K}$ | $m^{Mx}$ | $m^{My}$ | $m^{M\Gamma}$ |
| DFT | 1.16 | 1.99 | 1.46 | 0.97 | 1.27 | 0.32 | 0.31 | 0.77 | 0.31 |
| TB | 1.26 | 1.95 | 1.66 | 1.12 | 1.04 | 0.35 | 0.33 | 0.72 | 0.38 |

On the other hand, ML-SnS$_2$ is reported to have high on/off ratio in field effect transistors. As an application of the present effective Hamiltonian, we will focus on the tunable electronic structure via an external gate voltage $U$, which is introduced by setting the on-site potential on the two S-atom sublattices to different values:

$$\varepsilon_{mi} = \begin{cases} U/2, \text{if site } i \text{ on sublattice } A \\ -U/2, \text{if site } i \text{ on sublattice } B \end{cases}, (12)$$

In this case, only an unscreened electric field is considered, that is, we neglect explicit treatment of polarization and local-field effects. In other words, $U$ can be regarded as a local bias voltage assumed to be constant inside the sample and can be related to real external electric field $E_{ext}$ upon taking into account the thickness-dependent transverse dielectric permittivity and finite-size effect. And the equivalent electric field strength for $U$ = 1 eV is estimated around 3.12 eV/nm.

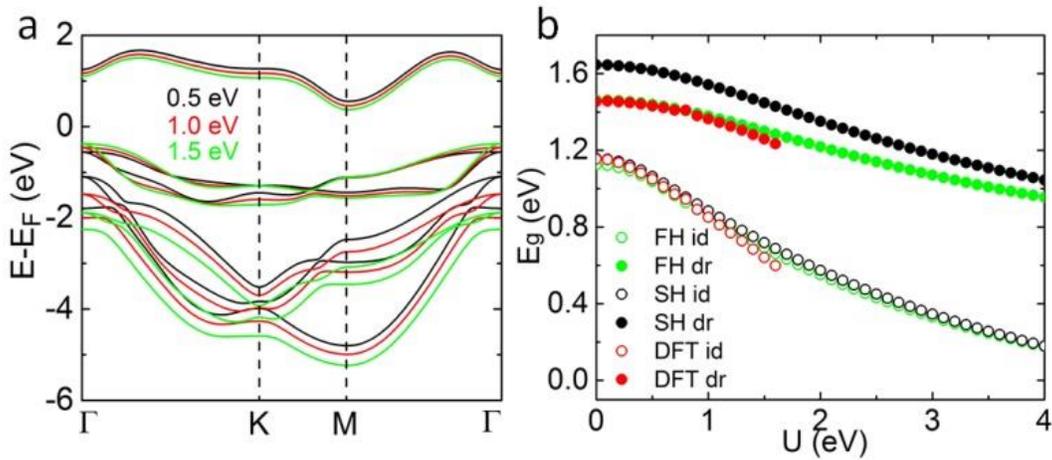

**Figure 5** Tunable electronic properties of ML-SnS$_2$ with external gate voltage. (a) Band structures of ML-SnS$_2$ calculated for different magnitudes of the gate voltage. (b) Band gap modulation of ML-SnS$_2$. The red, black and green symbols represent the DFT, simple Hamiltonian (SH) and full Hamiltonian (FH) results, respectively. And the solid and hollow symbols indicate the direct and indirect band gap, respectively.

The band gap modulation of ML-SnS$_2$ as a function of external gate voltage is plotted in Figure 5. As the bias voltage $U$ is applied, the interlayer potential increases, and the electronic bands shift due to the Stark effect. In Figure 5a, we show the band structure calculated for three representative gate voltages. Since the first CB is mainly composed of the *s* orbital of the central Sn atom, it is only slightly shifted with increasing gate voltage. On the contrary, the valence bands show remarkable changes (both shift and deformation) because the *p* orbitals of the top and bottom S atoms are the main contributors to these states, and they are directly affected by the on-site potential. Thus, for increasing gate voltage, the VB (and to a lesser extent, the CB) shifts toward the band gap center, and both the indirect and direct band gaps decrease as shown in Figure 5b. The band gap from DFT and full Hamiltonian (FH) calculations is also presented as a reference. The FH result coincides with DFT calculations in a wide energy window around 1 eV, because neither the orbital shape nor the hoppings are changed significantly by applying a small gate voltage. When much higher $U$ is applied, other effects like polarization must be taken into account to better describe the electronic properties. Our simple TB model agrees well with the FH model with gate voltage, especially for the indirect band gap variation between Γ and M. For the secondary direct band gap, there is good qualitative but no precise quantitative agreement. Of course, this is a consequence of the fact that only next-nearest-neighbor hoppings are taken into account in the simple TB model. Introducing more hoppings into the model would improve the direct band gap agreement, but would make the model more unwieldy.

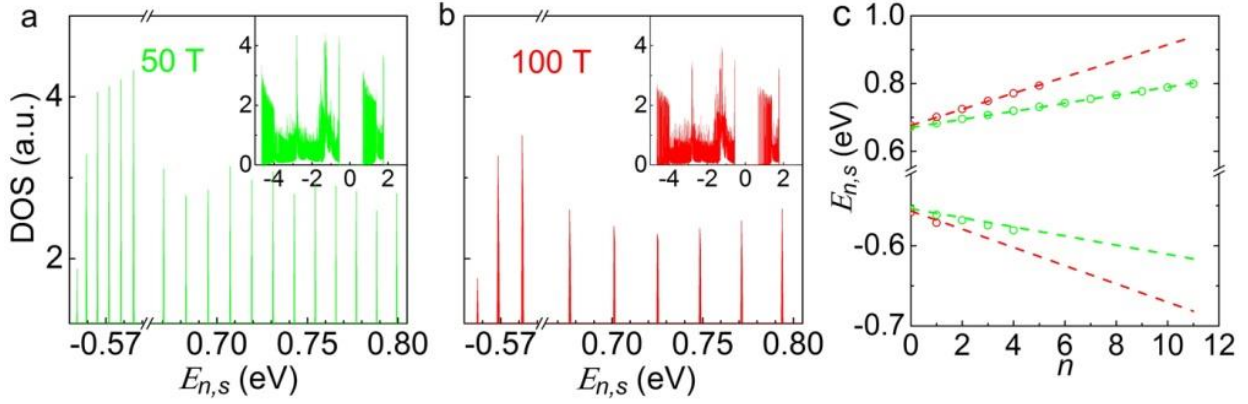

**Figure 6** Discrete Landau levels in ML-SnS$_2$. A super cell of ML-SnS$_2$ containing 4.344 million atoms (3*1200*1200) is simulated in the presence of a perpendicular magnetic field. The obtained DOS in the low energy region with magnetic fields **B** = 50 T and 100 T are indicated by green (a) and red (b) lines. The inset shows the density of states in a wider energy window. (c) Original (circles) and numerical fitting (dashed lines) of Landau levels in ML-SnS$_2$.

In the presence of a perpendicular magnetic field, the quantization of the energy eigenstates leads to discrete Landau levels (LLs) in many two-dimensional materials. [46-48] A clear splitting of the LLs is obtained in ML-SnS$_2$ for two different magnetic field strengths, as shown in Figure 6. As the energy at the conduction and valence band edge is almost parabolic with respect to $k$, the LLs at the low-energy region are linearly dispersed. Since the system lacks electron-hole symmetry, the cyclotron frequency is different for the valence and conduction bands. The obtained DOS consists of two sets of equidistant LLs described as $\varepsilon_{n,s}^{kp} = E_s + \frac{seB\hbar}{m_e}(n+1/2)\omega_s$, where $s = \pm 1$ denotes the conduction and valence bands, $E_{+/-} = E_{c/v}$ is the energy at the conduction and valence edge, $n$ is the energy index and $\omega_{+/-} = m_e/(m_x^{c,v} m_y^{c,v})^{1/2}$ (where $m_x^{c,v}$ and $m_y^{c,v}$ are the anisotropic effective masses at the conduction and valence edge obtained from Table 2). This is in good agreement with our numerical values of E$_{+/-}$ = 0.664 eV (-0.551 eV) and ω$_{+/-}$ = 2.052 (0.927) as indicated by the dashed line in Figure 6c.

**Conclusion**

To conclude, we have presented a systematic analysis on the electronic properties of ML-SnS$_2$. To this end, we performed relativistic first-principles calculations and derived a symmetry-based band TB model. We have shown that the *s* orbital of the Sn atom and the *p* orbitals of the S atom play a crucial role in determining the band structure of ML-SnS$_2$. We proposed an effective Hamiltonian, constructed with six hybrid *sp*-like orbitals, which shows good agreement with DFT calculations on electronic properties. Moreover, the proposed TB model is substantially less computationally demanding than first-principles calculations. This makes it suitable for a wide range of purposes, particularly for large-scale simulations of realistic ML-SnS$_2$ nano sheets and its heterostructures. We used the model for two applications. Firstly, we considered the case of applying a gate voltage, which reproduces the band gap modulation of ML-SnS$_2$ as full TB model and DFT results. Secondly, we studied the Landau levels in ML-SnS$_2$, which are equidistant, as expected for a two-dimensional electron gas.

## Author information

Correspondence should be addressed to j.yu@science.ru.nl and s.yuan@whu.edu.cn .

## Acknowledgements

Yu acknowledges financial support from MOST 2017YFA0303404, NSAF U1530401 and computational resources from the Beijing Computational Science Research Center. Katsnelson acknowledges financial support from the European Research Council Advanced Grant program (Contract No. 338957). Yuan acknowledges financial support from Thousand Young Talent Plan (China). The authors would like to thank Dr. Alexander Rudenko for his helpful discussions.